\documentstyle[aps,prb,english,theorem,makeidx,amsmath,epic,floats,
               amscd,curvesls,psfig,bbm,brakets,amssymb]{revtex}
\newtheorem{nota-rem}[defn]{Notation--Remark}

\def\sm{\sigma}
\def\bsm{\vec{\bf{\sigma}}}

\begin{document}
\title{A non-perturbative real-space renormalization group scheme for
       the spin-$\frac{1}{2}$ XXX Heisenberg model}
\author{Andreas Degenhard\thanks{e-mail: andreasd@icr.ac.de}}
\address{The Institute of Cancer Research, Department of Physics,\\
         Downs Road, Sutton, Surrey SM2 5PT, UK\\[0.6cm]}
\maketitle
\centerline{\bf{Abstract}}
\begin{abstract}
  In this article we apply a recently invented analytical real-space
  renormalization group formulation which is based on numerical
  concepts of the density matrix renormalization group. Within a
  rigorous mathematical framework we construct non-perturbative
  renormalization group transformations for the spin-$\frac{1}{2}$ XXX
  Heisenberg model in the finite temperature regime. The developed
  renormalization group scheme allows for calculating the renormalization
  group flow behaviour in the temperature dependent coupling constant.
  The constructed renormalization group transformations are applied
  within the ferromagnetic and the anti-ferromagnetic regime of the
  Heisenberg chain. The ferromagnetic fixed point is computed and
  compared to results derived by other techniques.  
\end{abstract}
\vskip 2mm
\medskip
~~~PACS. 75.10.Jm \;,\;\; PACS. 05.10.Cc
\vskip 8mm
%]
%
\section{Introduction}
In 1966 L.P. Kadanoff \cite{ka66} presented arguments which would allow
one to calculate critical exponents without ever working out the partition
function explicitly. Later in 1971 K.G. Wilson \cite{wi71} transformed
Kadanoff's qualitative block spin explanations into a quantitative
formulation in the context of critical phenomena which has become known
as the Wilson Renormalization Group (RG) technique.\\
The core physical concept of the RG is the scale invariance at the
critical point. As the physical system moves towards a phase transition,
it becomes increasingly dominated by large-scale fluctuations. At the
critical point the correlation length, i.e. the length scale of the
fluctuations, becomes infinite and the system exhibits scale invariance.
Within an application of the RG a RG transformation (RGT) needs to be
defined which eliminates non-relevant degrees of freedom of the system.
If the system becomes invariant by carrying out successive RG
transformations, a RG fixed point is reached. A statistical system
exhibits two trivial fixed points, belonging to zero and infinite
temperature, where the system possesses inherent scale invariance
according to complete order or complete disorder respectively. A
nontrivial fixed point, if present, corresponds to a critical
point of the physical system. The behaviour of physical quantities at
the critical point of the system is described by scaling laws using
critical exponents.\\
Although RG methods have been successfully applied to a variety of
physical problems, the construction of RGTs applicable to strong
coupling regimes is, apart from a few exceptions, an unsolved and
challenging problem. Examples of current research are the strong
coupling quantum spin chains, including the Heisenberg
models \cite{{gmdsv95},{so96}}, and nonlinear partial differential
equations (PDEs) \cite{{gmh98},{cmp98}}. Standard approaches like
perturbation theory in combination with Fourier space RG methods cannot
be applied.\\
In this paper we explore the critical behaviour of the isotropic
spin-$\frac{1}{2}$ Heisenberg model at finite temperature using the
Generalized Real-space Renormalization Group (GRRG). The method was
introduced in an earlier work as the general (real-space) RG\cite{de00}.
The GRRG requires the definition of an {\em auxiliary space}
${\cal H}_{\text{aux}}$ prior to the construction of the desired local
GRRG transformation (GRRGT), i.e. the RGT for a subsystem called a block,
analogous to the Kadanoff block spin formulation. The auxiliary space
describes the quantum correlations resulting from the different boundary
conditions between separated
adjacent blocks within the local GRRGT. If the defined auxiliary space
allows for a decomposition of the quantum system into blocks keeping all
possible boundary conditions the local RGT is called {\em exact}. In this
work we construct a {\em perfect} local GRRGT based on an auxiliary space
providing an approximate description of the boundary conditions. The local
GRRGT is formulated as a composition of two linear maps, the {\em embedding}
map $G$ and the {\em truncation} map $G^+$. Both maps depend on the choice
of the auxiliary space and are constructed according to an imposed physical
constraint, the {\em invariance relation} \cite{de00}. Originally the
conservation of the free energy or the partition function was used to
formulate the physical constraint.\cite{ka76}.\\
In this work we use the notation of the original work \cite{de00}. Block
quantities are indexed by capital letters, corresponding to sites in the
blocked chain. The indexing set for the blocks is denoted as ${\frak I}$.
Neighbouring blocks are indexed by a sequence
$I, I-1, I-2, \dots \in {\frak I}$ whereas independent blocks are indexed
by different letters $I, J, \dots \in {\frak I}$. A block Hilbert space
${\cal H}_I$ contains at minimum two single site Hilbert spaces
${\cal H}_i$ and ${\cal H}_{i-1}$. Single site Hilbert spaces are indexed
by letters $i,j,k,\dots$. If a single site space ${\cal H}_i$ is contained
within a block Hilbert space ${\cal H}_I$ we write
${\cal H}_i \subset{\cal H}_I$ or $i\in I$ if it is obvious that $I$ refers
to the block Hilbert space. We further use the abbreviation
$\{i,i-1,\dots\}\subset\{I,I-1,\dots\}$ instead of
${\cal H}_{\{i,i-1,\dots\}}\subset {\cal H}_{\{I\}}
\otimes{\cal H}_{\{I-1\}},\dots$. Using this notation it is not apparent
which single site space is contained in a particular block Hilbert space.
If this is important it needs to be pointed out explicitly.\\
In the next section we begin by revisiting the classical case. No quantum
correlations occur in the classical analogue of the Heisenberg chain
and the concept of an auxiliary space is therefore unnecessary. However,
we construct a local RGT and by direct comparison with the GRRG method
we provide the reader with a feel for the abstract mathematical formulation
of the GRRG method. We proceed in section \ref{quancorrt} by constructing a
perfect local RGT for the spin-$\frac{1}{2}$ XXX Heisenberg model following
the concepts in the original work \cite{de00}. In section \ref{quanflowt}
we discuss analytical results of the flow behaviour calculated by the
perfect GRRGT. Critical exponents are calculated from the nontrivial
ferromagnetic RG flow behaviour for the three dimensional Heisenberg
chain and compared with results calculated by other methods. In
section \ref{quancorrtanti} we derive the flow behaviour for the
anti-ferromagnetic regime of the one-dimensional Heisenberg model. In the
final section we conclude with some perspectives on the GRRG method.
%
%%%%%%%%%%%%%%%%%%%%%%%%%%%%%%%%%%%%%%%%%%%%%%%%%%%%%
\section{The Migdal-Kadanoff RGT}
\label{ising}
%%%%%%%%%%%%%%%%%%%%%%%%%%%%%%%%%%%%%%%%%%%%%%%%%%%%%
%
We consider the one dimensional Ising model without an external
magnetic field and with nearest neighbour (n.n.) interaction \cite{hk73}.
All thermodynamic quantities can be calculated from the model's partition
function
\begin{align}
\label{mig1}
 {\cal Z_{\text{Ising}}}\,=\,\sum_{\{\sm_j\}}\,e^{-1/({k_B\, T})\,H_{\text{Ising}}}
  \qquad\text{with}\quad H_{\text{Ising}}\,=\,\sum_{i=1}^N\, J\,\sm_i\,\sm_{i+1}\;,
\end{align}
where $\{\sm_i\}$ denotes that the sum should be extended over all
possible assignments of $\pm 1$ to each lattice site $i$ corresponding to
an array of elementary spins $\{\sm_i\}$ placed on the lattice sites
$\{i\}$. In (\ref{mig1}) $k_B$ is Boltzmann's constant and we are
interested in the limit $N\longrightarrow \infty$.\\
Introducing a temperature dependent coupling constant
\begin{align}
\label{mig2}
 K\,=\,\frac{-J}{k_B\, T}
\end{align}
we write the partition function in the form
\begin{align}
\label{mig3}
{\cal Z_{\text{Ising}}}\,=\,\sum_{\{\sm_j\}}\,\prod_{i}^{N/2}\,
 \exp\left[K\left(\sm_{2i-1}\sm_{2i}+\sm_{2i}\sm_{2i+1}\right)\right]\,.
\end{align}
Decomposing the sum over all possible configurations into odd and even
sites the sum over the even sites is calculated by successive application
of
\begin{align}
\label{loctr}
 \sum_{\sm_{2i}=\pm 1}\exp\left[K\left(\sm_{2i-1}\sm_{2i}+\sm_{2i}
  \sm_{2i+1}\right)\right]\,=\,
   \exp\left[K\left(\sm_{2i-1}+\sm_{2i+1}\right)\right]\,+\,
    \exp\left[-K\left(\sm_{2i-1}+\sm_{2i+1}\right)\right]
\end{align}
for every even site $2i$.
The application of the GRRG method requires a physical constraint to
construct a local GRRGT imposed by an {\em invariance relation} \cite{de00}.
We define the classical analogue of the invariance relation by keeping the
partition function unchanged
\begin{align}
\label{invar}
 {\cal Z}\left[H_{\text{Ising}}(J)\right]\,=\,
   {\cal Z}\left[H_{\text{Ising}}(J',f')\right]\;.
\end{align}
In equation (\ref{invar}) $f'$ denotes a change in the ground-state
energy of the energy function $H_{\text{Ising}}(J)$. By inserting
(\ref{mig1}) and (\ref{loctr}) into equation (\ref{invar}) we define
the effective functional dependence as
\begin{align}
\label{effobs}
 {\cal O}(\{\sm_j\},K',f')\,=\,
  \prod_{i}^{N/4}\,
   \exp\left[K'\left(\sm_{2i-1}\sm_{2i}+\sm_{2i}\sm_{2i+1}\right)\,+\,4f'/N\right]\;.
\end{align}
Using the definition (\ref{effobs}) we derive the classical analogue
of the global GRRGT as
\begin{align}
\label{globclassic}
  \sum_{\{\sm_j,\;j\,\text{even}\}}\,\prod_{i}^{N/2}\,
   \exp\left[K\left(\sm_{2i-1}\sm_{2i}+\sm_{2i}\sm_{2i+1}\right)\right]
  &\;\,=\,\prod_{i}^{N/4}\,
   \exp\left[K'\left(\sm_{2i-1}\sm_{2i}+\sm_{2i}\sm_{2i+1}\right)\,+\,4f'/N\right]\;.
\end{align}
The GRRG method requires a decomposition of the spin chain into commuting
blocks, which can always be performed in the classical case. We therefore
write the local GRRGT as
\begin{align}
\label{invarre}
 \sum_{\sm_{2i}=\pm 1}\exp\left[K\left(\sm_{2i-1}\sm_{2i}+\sm_{2i}
     \sm_{2i+1}\right)\right]\,=\,
      \exp\left[K'\,\sm_{2i-1}\,\sm_{2i+1}\,+\,4f'/N\right]
\end{align}
and the effective coupling is calculated as $K' =(1/2)\cdot\ln\cosh(2K)$
which yields to the trivial RG flow behaviour as it is expected for one
dimensional strongly correlated systems \cite{mw66}. Relation
(\ref{invarre}) is the classical analogue of an {\em exact} local GRRGT
since the invariance relation (\ref{invar}) can be derived from the local
GRRGT.\\
The previous calculations are a reinterpretation of the Migdal-Kadanoff
transformation for classical spin systems. The calculation was first
done by A.A. Migdal \cite{{mi76a},{mi76b}} and reformulated using bond
moving techniques by L.P. Kadanoff \cite{{hk75},{ka76}}.
%
%%%%%%%%%%%%%%%%%%%%%%%%%%%%%%%%%%%%%%%%%%%%%%%%%%%%%
\section{The construction of the local RGT for the isotropic Heisenberg chain}
\label{quancorrt}
\begin{figure}[ht]
\vspace{0.6cm}
\centerline{
  \psfig{figure=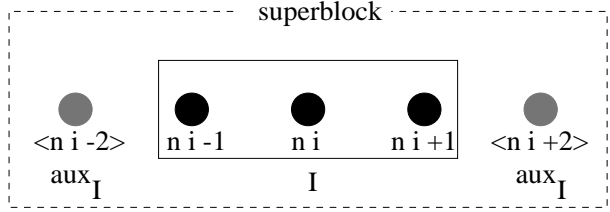,width=8cm,height=2.8cm}}
\vspace{0.8cm}
\caption{\em A system block enlarged by copies of the two nearest
  neighbour sites. The additional sites represent the auxiliary space
  and all single site Hilbert spaces together define the superblock.}
  \label{neighbour}
\end{figure}

\begin{figure}[ht]
\vspace{0.6cm}
\centerline{
  \psfig{figure=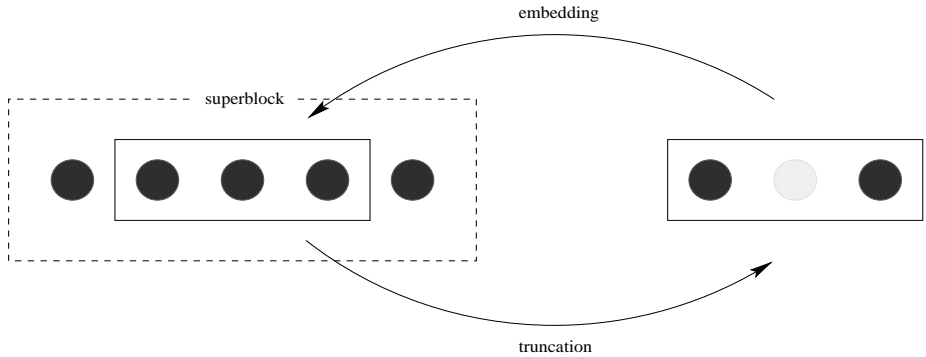,width=12.2cm,height=4.6cm}}
\vspace{0.8cm}
\caption{\em The local embedding $G_{{\cal H}'\otimes{\cal H}'_{\text{aux}}}$
  and truncation $G^+_{{\cal H}\otimes{\cal H}_{\text{aux}}}$ procedure within
  the superblock. The lightly shaded dot is truncated within the spin
  decimation of the system block.}
  \label{embtruncsup}
\end{figure}

%%%%%%%%%%%%%%%%%%%%%%%%%%%%%%%%%%%%%%%%%%%%%%%%%%%%%
%
In this section we derive a perfect local RGT for the isotropic
spin-$\frac{1}{2}$ Heisenberg chain by applying the GRRG method \cite{de00}.
To the best knowledge of the author no other controllable approximation
is currently available to analytically calculate the critical properties
for the quantum spin chain at finite temperature.\\
The Hamiltonian of the model is defined by
\begin{align}
\label{heisb1}
 H_{XXX}\,=\, J\sum_{i=1}^N\left(\sm^x_i\sm^x_{i+1}\,+\,\sm^y_i\sm^y_{i+1}
                  \,+\,\sm^z_i\sm^z_{i+1}\right)\,,
\end{align}
which is totally isotropic in the spin components and known as the
XXX spin-$\frac{1}{2}$ model \cite{{he28},{bl30}}. The spin variables
$\sm^x$ , $\sm^y$ and $\sm^z$ define the Lie algebra ${\frak
  {sl}}(2)$.
In this article we choose the smallest nontrivial representation
$S_i^{\alpha}=(\hbar/2)\sm^{\alpha}_i$ by the Pauli matrices\\
\begin{align}
\label{heisb5}
 \sm^x=\begin{pmatrix}0 & 1 \\ 1 & 0 \end{pmatrix}\,,\qquad
 \sm^y=\begin{pmatrix}0 & -i \\ i & 0 \end{pmatrix}\,,\qquad
 \text{and}\qquad\sm^z=\begin{pmatrix}1 & 0 \\ 0 & -1 \end{pmatrix}\;.
\end{align}
The partition function for the one-dimensional Heisenberg model is
defined by\\
\begin{align}
\label{partheisb1}
 {\cal Z_{\text{XXX}}}
  \,=\, &\text{tr}_{\{\sm_j\}}\,\exp\left[\frac{-J}{k_B\, T}\,
   \sum_{i=1}^N\left(\sm^x_i\sm^x_{i+1}\,+\,\sm^y_i\sm^y_{i+1}
                  \,+\,\sm^z_i\sm^z_{i+1}\right)\right]\nonumber\\[0.2cm]
  \,=\, &\text{tr}_{\{\sm_j\}}\,\exp\left[K\,\sum_{i=1}^N\,\bsm_i\cdot\bsm_{i+1}
        \right]
\end{align}
where we used the vector notation and introduced a temperature dependent coupling
constant $K\,=\,\frac{-J}{k_B\, T}$. In equation (\ref{partheisb1})
$\text{tr}_{\{\sm_j\}}$ denotes the trace over all lattice sites in the quantum
chain. Analogous to the classical case reported in section \ref{ising} the
partition function is used for defining the invariance relation
\begin{align}
\label{partheisb2}
{\cal Z_{\text{XXX}}}\left[{\cal O}\left(\left\{\sm_j\right\},K\right)\right]
 \,=\, &\,\text{tr}_{\{\sm_j,\;j\,\text{odd}\}}\,\text{tr}_{\{\sm_j,\;j\,
     \text{even}\}}\,\exp\left[K\,\sum_{i=1}^N\,\bsm_i\cdot\bsm_{i+1}
       \right]\nonumber\\[0.2cm]
  \,:=\, &\,\text{tr}_{\{\sm_j,\;j\,\text{odd}\}}\,
      \exp\left[f'(K)\,+\,K'(K)\,\sum_{i=1}^{N/2}\,\bsm_{2i-1}\cdot\bsm_{2i+1}
          \right]\nonumber\\[0.2cm]
          \,=\, &\,{\cal
 Z_{\text{XXX}}}\left[{\cal O}\left(\left\{\sm_j,\;j\,\text{odd}\right\}
       ,K',f'\right)\right]\nonumber\\
\end{align}
where we used of the factorization property of the trace
$\text{tr}_{\{\sm_j\}}=\prod_i\,\text{tr}_{\sm_i}$.\\
According to the construction of the local GRRGT we proceed by changing our
notation and equip every operator with an abstract auxiliary space \cite{de00}
which is currently not further specified. The action of each of the operators
on the auxiliary space is defined as the identity map until further
specifications are given. The embedding map
$G_{{\cal H}'\otimes{\cal H}'_{\text{aux}}}$ and the truncation map
$G^+_{{\cal H}\otimes{\cal H}_{\text{aux}}}$ are defined according to the
invariance relation
\begin{equation}
\label{inveq2}
{\cal Z_{\text{XXX}}}\left[{\cal O}_{{\cal H}\otimes{\cal H}_{\text{aux}}}
  \left({\bf K}\right)\right]\,=\,{\cal Z_{\text{XXX}}}\left[
  G^+_{{\cal H}\otimes{\cal H}_{\text{aux}}}\,\circ\,
  {\cal O}_{{\cal H}\otimes{\cal H}_{\text{aux}}}\left({\bf K}\right)\,\circ\,
  G_{{\cal H}\otimes{\cal H}_{\text{aux}}}\right]
\,=\,{\cal Z_{\text{XXX}}}\left[{\cal O}_{{\cal H}'\otimes{\cal H}'_{\text{aux}}}
  \left({\bf K}'\right)\right]\,.
\end{equation}
The embedding map $G_{{\cal H}'\otimes{\cal H}'_{\text{aux}}}$ together with
the truncation map $G^+_{{\cal H}\otimes{\cal H}_{\text{aux}}}$ define the GRRGT.
By choosing
\begin{equation}
\label{choice1}
 {\cal O}_{{\cal H}\otimes{\cal H}_{\text{aux}}}\left({\bf K}\right)\,=\,
 \exp\left[K\,\sum_{i=1}^N\,\bsm_i\cdot\bsm_{i+1}\right]
 \qquad\text{with}\quad {\bf K}=(K,f)
\end{equation}
the embedding and truncation maps are defined as
\begin{equation}
\label{truncemb}
 G^+_{{\cal H}\otimes{\cal H}_{\text{aux}}}\,=\,
 \text{tr}_{\{\sm_j,\;j\,\text{even}\}}\otimes
  {\mathbbm{1}}_{\{\sm_j,\;j\,\text{odd}\}}\otimes
   {\mathbbm{1}}_{{\cal H}_{\text{aux}}}\qquad\text{and}\qquad
   G_{{\cal H}'\otimes{\cal H}'_{\text{aux}}}\,=\,
    {\mathbbm{1}}_{{\cal H}'\otimes{\cal H}'_{\text{aux}}}\;.
\end{equation}
Analogous to the classical case the local operators are derived from their
global counterparts by decomposing the quantum chain into blocks. Taking
the trace over one even site in each block we arrive at the relation
\begin{align}
\label{partheisb3}
{\cal Z_{\text{XXX}}} &\left[{\cal O}_{{\cal H} \otimes{\cal H}_{\text{aux}}}
  \left({\bf K}\right)\right]
 \,=\,\text{tr}_{\{\sm_j,\;j\,\text{odd}\}}\,\text{tr}_{\{\sm_j,\;j\,
  \text{even}\}}\,\exp\left[K\,\sum_{i=1}^N\,\bsm_i\cdot\bsm_{i+1}
   \right]\,\otimes{\mathbbm{1}}_{{\cal H}_{\text{aux}}}\nonumber\\[0.2cm]
 \,= &\;\text{tr}_{\{\sm_j,\;j\,\text{odd}\}}\,\prod_{I=1}^{N/2}\,
 \text{tr}_{\text{even}}\Big\{\,\exp\left[K\,H_{{\cal H}_{I}}\right]
  \,\otimes{\mathbbm{1}}_{{\cal H}_{\text{aux}}}
\otimes\,{\cal C}_{{\cal H}_{I}\otimes{\cal H}_{I+1}}
  \!\left({\bf K}\right)\,\Big\}\nonumber\\[0.2cm]
 \,= &\;\text{tr}_{\{\sm_j,\;j\,\text{odd}\}}\,\prod_{I=1}^{N/2}\,
 G^+_{{\cal H}_{I}\,\otimes\,
   {\left({\cal H}_{\text{aux}}\right)}_{I}}
    \Big\{\,{{\cal O}^{\text{system}}_{{\cal H}_{I}\otimes
         {\left({\cal H}_{\text{aux}}\right)}_{I}}
      \!\left({\bf K}\right)}
   \otimes\,{{\cal O}^{\text{correlation}}_{{\cal H}_{I}
    \otimes {\cal H}_{I+1}\otimes
         {\left({\cal H}_{\text{aux}}\right)}_{I,I+1}}
      \!\left({\bf K}\right)}\,\Big\}\nonumber\\[-0.2cm]
\end{align}
containing only local operators for block Hilbert spaces. In equation
(\ref{partheisb3}) we identified
$H_{{\cal H}_{I}}=\bsm_{2i-1}\cdot\bsm_{2i}\,+\,\bsm_{2i}\cdot\bsm_{2i+1}$
to separate the dependence on the parameter $K$. The block decomposition of the
functional dependence ${\cal O}$ in (\ref{partheisb3}) contains two parts
defined as
\begin{align}
\label{sysdef}
{{\cal O}^{\text{system}}_{{\cal H}_{I}\otimes
  {\left({\cal H}_{\text{aux}}\right)}_{I}}\!\left({\bf K}\right)} \,=\,
    \exp\left[K\,H_{{\cal H}_{I}}\right]
\end{align}
and
\begin{align}
\label{corrdef}
{{\cal O}^{\text{correlation}}_{{\cal H}_{I}\otimes {\cal H}_{I+1}\otimes
    {\left({\cal H}_{\text{aux}}\right)}_{I,I+1}}\!\left({\bf K}\right)}\,=\,
  {\cal C}_{{\cal H}_{I}\otimes{\cal H}_{I+1}}\left({\bf K}\right)
\end{align}
following the nomenclature for the block decomposition in the original work
\cite{de00}. Ignoring the {\em correlation block part} (\ref{corrdef})
in (\ref{partheisb3}) yields a decomposition of the quantum chain into
commuting {\em system block operators} (\ref{sysdef}) analogous to the
classical situation. This approximation is valid in the high temperature
limit $T\longrightarrow\infty$ where higher order terms of the coupling $K$,
included in the correlation block part, vanish.\\
The correlation block part (\ref{corrdef}) is calculated
using the Baker-Campbell-Hausdorf formula \cite{su76} for blocks
\begin{align*}
 \exp\left[K \left(\,H_{{\cal H}_{I}}\,+\, H_{{\cal H}_{I+1}}\,\right)\right]
  =\,\exp\left(K\,H_{{\cal H}_{I}}\right)\,
  \exp\left(K\,H_{{\cal H}_{I+1}}\right)\,\cdot\,{\cal C}_{{\cal H}_{I}
    \otimes{\cal H}_{I+1}}\left({\bf K}\right)\\[0.2cm]
  \text{with}\qquad {\cal C}_{{\cal H}_{I}\otimes{\cal H}_{I+1}}
   \left({\bf K}\right)\,= \,
    \exp\left(\frac{K^2}{2}\,\left[H_{{\cal H}_{I}},\,H_{{\cal H}_{I+1}}\right]
     \,+\,\dots\right)\,.
\end{align*}
Relation (\ref{partheisb3}) is an example for a {\em product block decomposition}
\cite{de00}. The separated correlation block part
${\cal C}_{{\cal H}_{I}\otimes{\cal H}_{I+1}}\left({\bf K}\right)$ of the
functional dependence describes
the quantum correlations between adjacent blocks in the decomposition. To
eliminate the correlation block part (\ref{corrdef}) in relation
(\ref{partheisb3}) an auxiliary space needs to be defined to describe the
boundary conditions between adjacent blocks. Dependent on the choice of the
auxiliary space ${\left({\cal H}_{\text{aux}}\right)}_{I}$ the action of the
block operators on the auxiliary space is determined.\\
To describe the quantum correlations between the block $I$ and the
neighbouring blocks $I-1$ and $I+1$ the correlation block operator
(\ref{corrdef}) includes the coupling between the nearest neighbour (n.n.)
single site spins of adjacent blocks. To include this n.n. coupling into
the description of the boundary conditions we construct an auxiliary space
including the n.n. spin sites of the block Hilbert space by choosing copies
of the n.n. sites of the system block as visualized in figure \ref{neighbour}.
% %%%%%%%%%%%%%%%%%%%%%%%%%%%%%%%%%%%%%%%%%%%%%%%%%%%%%%%%%%%%%%%%%
%
% One dimensional auxspace visualization.
%
% %%%%%%%%%%%%%%%%%%%%%%%%%%%%%%%%%%%%%%%%%%%%%%%%%%%%%%%%%%%%%%%%%
% %%%%%%%%%%%%%%%%%%%%%%%%%%%%%%%%%%%%%%%%%%%%%%%%%%%%%%%%%%%%%%%%%
According to the nomenclature in the original work \cite{de00} we call
the constructed space shown in figure \ref{neighbour} a {\em superblock}.
To distinguish between the original system block sites and the copies of
the n.n. sites in the superblock we mark the site indices of the copies
with brackets '$<.>$'.\\
Using the constructed auxiliary space together with relation
(\ref{partheisb3}) results in the approximation
\begin{align}
\label{finalheisb3}
{\cal Z_{\text{XXX}}} \left[{\cal O}_{{\cal H} \otimes{\cal H}_{\text{aux}}}
  \left({\bf K}\right)\right]
 \,\approx \,\text{tr}_{\{\sm_j,\;j\,\text{odd}\}}\,\prod_{I=1}^{N/2}\,
 G^+_{{\cal H}_{I}\,\otimes\,
   {\left({\cal H}_{\text{aux}}\right)}_{I}}
    \Big\{\,{{\cal O}^{\text{system}}_{{\cal H}_{I}\otimes
         {\left({\cal H}_{\text{aux}}\right)}_{I}}
      \!\left({\bf K}\right)}\,\Big\}\,.
\end{align}
Unlike in the classical case of section \ref{ising} the block decomposition
in relation (\ref{finalheisb3}) does not allow for an exact conservation of
the partition function. Furthermore, according to relation (\ref{finalheisb3}),
both auxiliary sites need to be truncated within the GRRGT and by the choice
of ${\cal H}_{\text{aux}}$ we define an example for an
{\em active auxiliary space}\cite{de00}.\\
Here we give two remarks on the foregoing calculation: The choice of the
particular auxiliary space allows for describing the boundary conditions
between the blocks which determine the quantum correlations. The
auxiliary space contains copies of the n.n. sites and neglects the effect
of next nearest neighbour and further higher order couplings.
Secondly we need to ensure that the auxiliary sites are treated as copies
of the original sites during the GRRGT. Otherwise only the block Hilbert
space ${\cal H}_I$ would have been enlarged and the description of the
boundary conditions will fail. Since the single site Hilbert spaces and
their copies are formally indistinguishable the identification as
auxiliary sites in the superblock
${{\cal H}_{I}\otimes{\left({\cal H}_{\text{aux}}\right)}_{I}}$ is
accomplished by the embedding and truncation operators
$G_{{\cal H}'_{I}\,\otimes\,{\left({\cal H}'_{\text{aux}}\right)}_{I}}$ and
$G^+_{{\cal H}_{I}\,\otimes\,{\left({\cal H}_{\text{aux}}\right)}_{I}}$
as illustrated in figure \ref{embtruncsup}.
% %%%%%%%%%%%%%%%%%%%%%%%%%%%%%%%%%%%%%%%%%%%%%%%%%%%%%%%%%%%%%%%%%
%
% Embedding and truncation within the superblock.
%
% %%%%%%%%%%%%%%%%%%%%%%%%%%%%%%%%%%%%%%%%%%%%%%%%%%%%%%%%%%%%%%%%%
% %%%%%%%%%%%%%%%%%%%%%%%%%%%%%%%%%%%%%%%%%%%%%%%%%%%%%%%%%%%%%%%%%
On the right hand side of figure \ref{embtruncsup} the system block is
visualized as the effective Hilbert space ${\cal H}'$ and the lightly
shaded point denotes an even site which has been truncated in the
GRRGT.\\
After the necessary definitions within the GRRG approach have been performed
the local RGT is given by the commuting diagram \cite{de00}
\begin{equation}
\label{comm4}
\begin{CD}
  {\cal H}'_{I} @> G_{{\cal H}'_{I}} >>
  {\cal H}_{I}\otimes {\left({\cal H}_{\text{aux}}\right)}_{I}\\ @V {{\cal
      O}_{{\cal H}'_{I}}
    \left({\bf K}'\right)} VV @VV {{\cal O}_{{\cal H}_{I}\otimes{\left({\cal
            H}_{\text{aux}}\right)}_{I}} \left({\bf K}\right)} V \\
  {\cal H}'_{I} @<< G^+_{{\cal
        H}_{I}\,\otimes\,{\left({\cal H}_{\text{aux}}\right)}_{I}} <
    {\cal H}_{I}\otimes {\left({\cal H}_{\text{aux}}\right)}_{I}\\
\end{CD}
\end{equation}
Due to the active auxiliary space no auxiliary sites are left after
applying the GRRGT. Within a successive application of the GRRGT new copies of
the changed n.n. sites for a system block need to be generated.\\
We summarize the local operators as
\begin{align}
 \label{summarize}
 {{\cal O}_{{\cal H}_{I}\otimes{\left({\cal H}_{\text{aux}}\right)}_{I}}}
   \left({\bf K}\right)
   \,= &\,
     \exp\left[K{H_{{\cal H}_{I}\otimes{\left({\cal H}_{\text{aux}}\right)}_{I}}}
       \right]\nonumber \allowdisplaybreaks\\[0.2cm]
     = &\,\exp\left[K\left({\bsm_{\langle 2i-2\rangle }\cdot\bsm_{2i-1}
       +\bsm_{2i-1}\cdot\bsm_{2i}+\bsm_{2i}\cdot\bsm_{2i+1}
      +\bsm_{2i+1}\cdot\bsm_{\langle 2i+2\rangle }}\right)\right]\,,
        \nonumber \allowdisplaybreaks\\[0.2cm]
 {\cal O}_{{\cal H}'_{I}}\left({\bf K}'\right)
  \,= &\,\exp\left[K'{H_{{\cal H}'_{I}}\,+\,f'}\right]
     \nonumber \allowdisplaybreaks\\[0.2cm]
   \,= &\,\exp\left[K'\,\left(\bsm_{2i-1}\cdot\bsm_{2i+1}\right)\,+\,f'\right]
  \qquad\quad\,\;\mbox{with}\quad i\in\{1,\dots ,N/2\}\quad\mbox{and}
      \nonumber \allowdisplaybreaks\\[0.4cm]
   G_{{\cal H}'_{I}}\,= &\,{\mathbbm{1}}_{{\cal H}'_{I}}\,,
      \nonumber \allowdisplaybreaks\\[0.2cm]
   G^+_{{\cal H}_{I}\,\otimes\,{\left({\cal H}_{\text{aux}}\right)}_{I}}
     \,= &\,\text{tr}_{\{\sm_i,\;i\,\text{even}\}}\otimes
       {\mathbbm{1}}_{\{\sm_i,\;i\,\text{odd}\}}\otimes
   \text{tr}_{{\cal H}_{\text{aux}}}
    \qquad\mbox{with}\quad i\in\{1,\dots ,N\}\;.
      \nonumber \allowdisplaybreaks\\[-0.2cm]
  \end{align}
By using (\ref{summarize}) and calculating
\begin{align}
 \label{qtransini}
  \prod_{I\in {\frak I}}{\cal O}_{{\cal H}'_{I}
  \otimes{\left({\cal H}'_{\text{aux}}\right)}_{I}}\left({\bf K}'\right)
 \,=\, &\,
 \prod_{I\in {\frak I}}\left\{\,
 \left[G^+_{{\cal H}_{I}\otimes{\left({\cal H}_{\text{aux}}\right)}_{I}}\right]
   \,\circ\, {{\cal O}^{\text{system}}_{{\cal H}_{I}\otimes
   {\left({\cal H}_{\text{aux}}\right)}_{I}} \!\left({\bf K}\right)}\,\circ\,
  \left[G_{{\cal H}'_{I}\otimes{\left({\cal H}'_{\text{aux}}\right)}_{I}}\right]
   \,\right\}\nonumber \allowdisplaybreaks\\[0.2cm]
   \,=\, &\,
  \text{tr}_{\text{even}}\,\prod_{I\in {\frak I}}\text{tr}_{\langle 2i-2\rangle ,\langle 2i+2\rangle }\,
  \exp\left[K\, H_{{\cal H}_{I}\otimes{\left({\cal H}_{\text{aux}}\right)}_{I}}
  \right]\,\nonumber \allowdisplaybreaks\\[0.2cm]
   \,=\, &\,
   \text{tr}_{\text{even}}\,\text{tr}^{{a}}_{\langle \text{even}\rangle }\,
   \text{tr}^{{b}}_{\langle \text{even}\rangle }\,
   \prod_{I\in {\frak I}}\,
   \exp\left[K\, H_{{\cal H}_{I}\otimes{\left({\cal H}_{\text{aux}}\right)}_{I}}
     \right]\,\nonumber\\[-0.2cm]
\end{align}
we proved that the GRRGT is perfect \cite{de00}. In the
final equation of (\ref{qtransini}) we have to trace over two copies of the
even sites denoted as $\text{tr}^{{a}}$ and $\text{tr}^{{b}}$.\\
The choice of the auxiliary space does not allow for an
exact treatment of the quantum correlations during the local RG procedure.
Our approach therefore yields a perfect instead of an exact GRRGT.
Although the product decomposition (\ref{partheisb3}) allows for further
improvement in the description of quantum correlations by increasing the
number of copied neighbouring sites it is not possible to construct
an exact GRRGT. The definition of an exact GRRGT is possible but requires
a different and more abstract auxiliary space. Results on the exact GRRGT
will therefore be reported elsewhere \cite{de01}. 
%
%%%%%%%%%%%%%%%%%%%%%%%%%%%%%%%%%%%%%%%%%%%%%%%%%%%%%
\section{The GRRG flow behaviour in the ferromagnetic regime}
\label{quanflowt}
\begin{figure}[ht]
\vspace{0.6cm}
\centerline{
  \psfig{figure=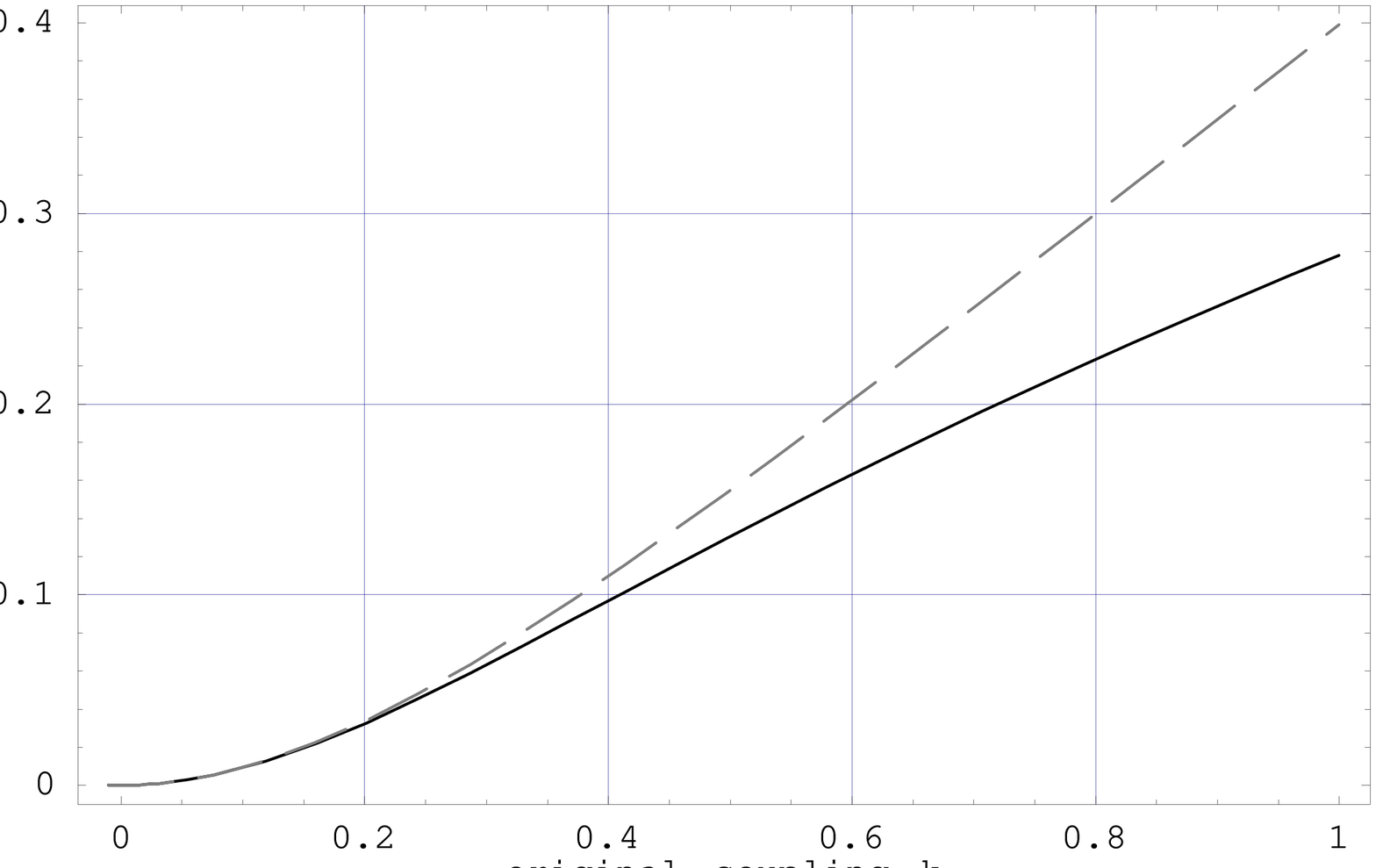,width=8cm,height=5.8cm}}
\vspace{0.8cm}
\caption{\em The calculated GRRG flow for the superblock in the
  ferromagnetic regime (solid curve). The dashed curve displays a RG
  flow using fixed boundary conditions for the system blocks, only
  valid in the high temperature limit.}
  \label{ferr}
\end{figure}

\begin{figure}[ht]
\vspace{0.6cm}
\centerline{
  \psfig{figure=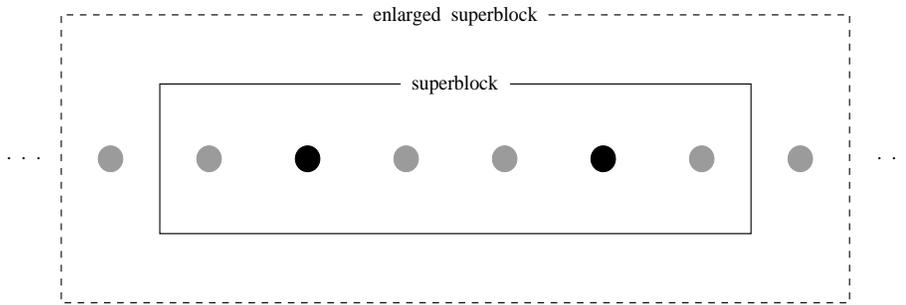,width=12.0cm,height=4.0cm}}
\vspace{0.8cm}
\caption{\em A superblock and an enlarged superblock construction for a
  GRRGT with a reduction factor $\lambda =3$. The lightly shaded dots
  are decimated in the local GRRGT.}
  \label{foursuper}
\end{figure}

\begin{figure}[ht]
\vspace{0.2cm}
\centerline{
  \psfig{figure=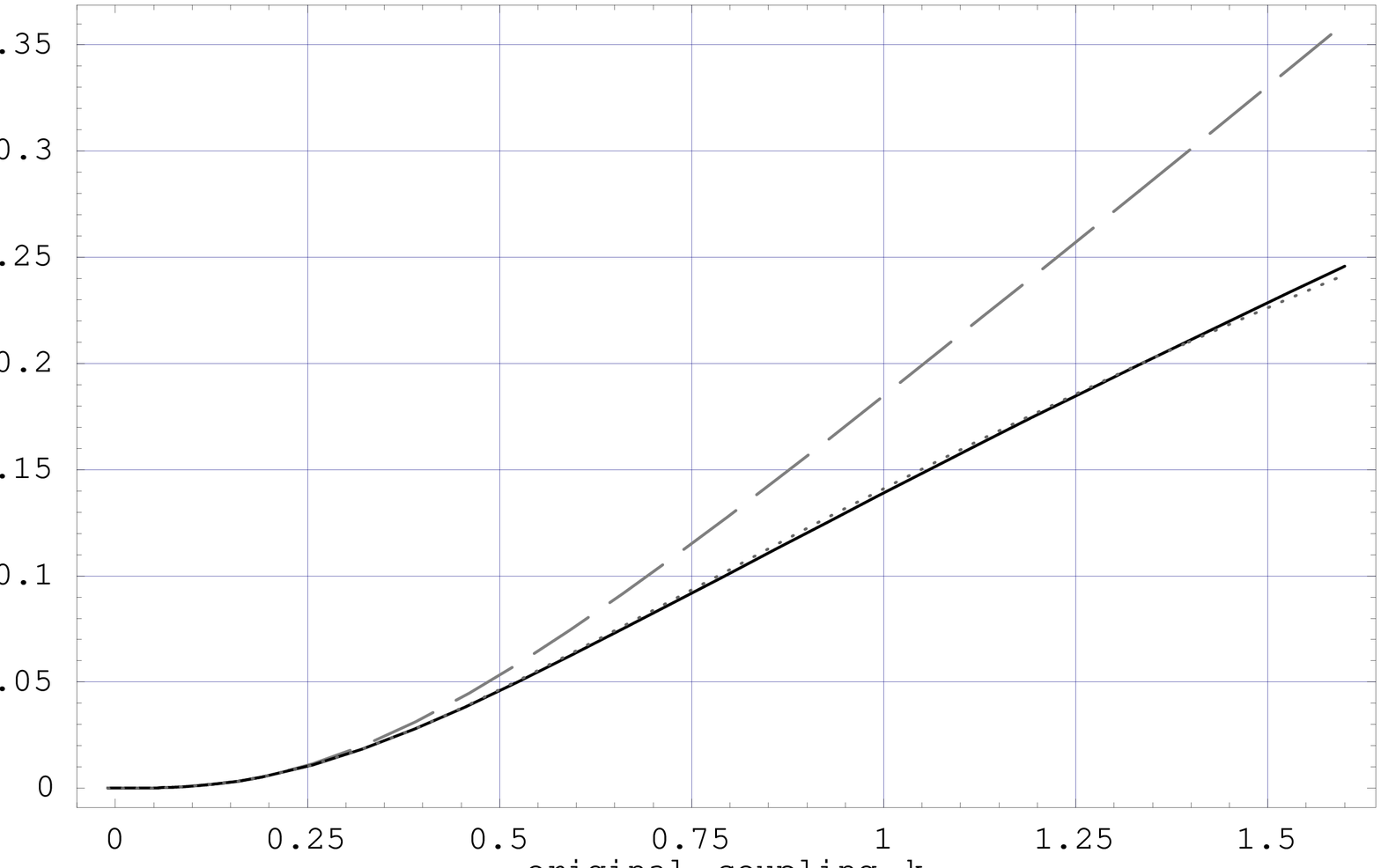,width=8cm,height=5.8cm}}
\vspace{0.8cm}
\caption{\em The superblock RG flow and the enlarged superblock
  RG flow in the ferromagnetic regime. The dotted curve shows the
  flow behaviour using the enlarged superblock as compared to the
  usual superblock construction (solid curve). The dashed curve
  displays a RG flow using fixed boundary conditions for the
  system blocks, only valid in the high temperature limit.}
  \label{fourplot}
\end{figure}

\begin{figure}[ht]
\vspace{0.2cm}
\centerline{
  \psfig{figure=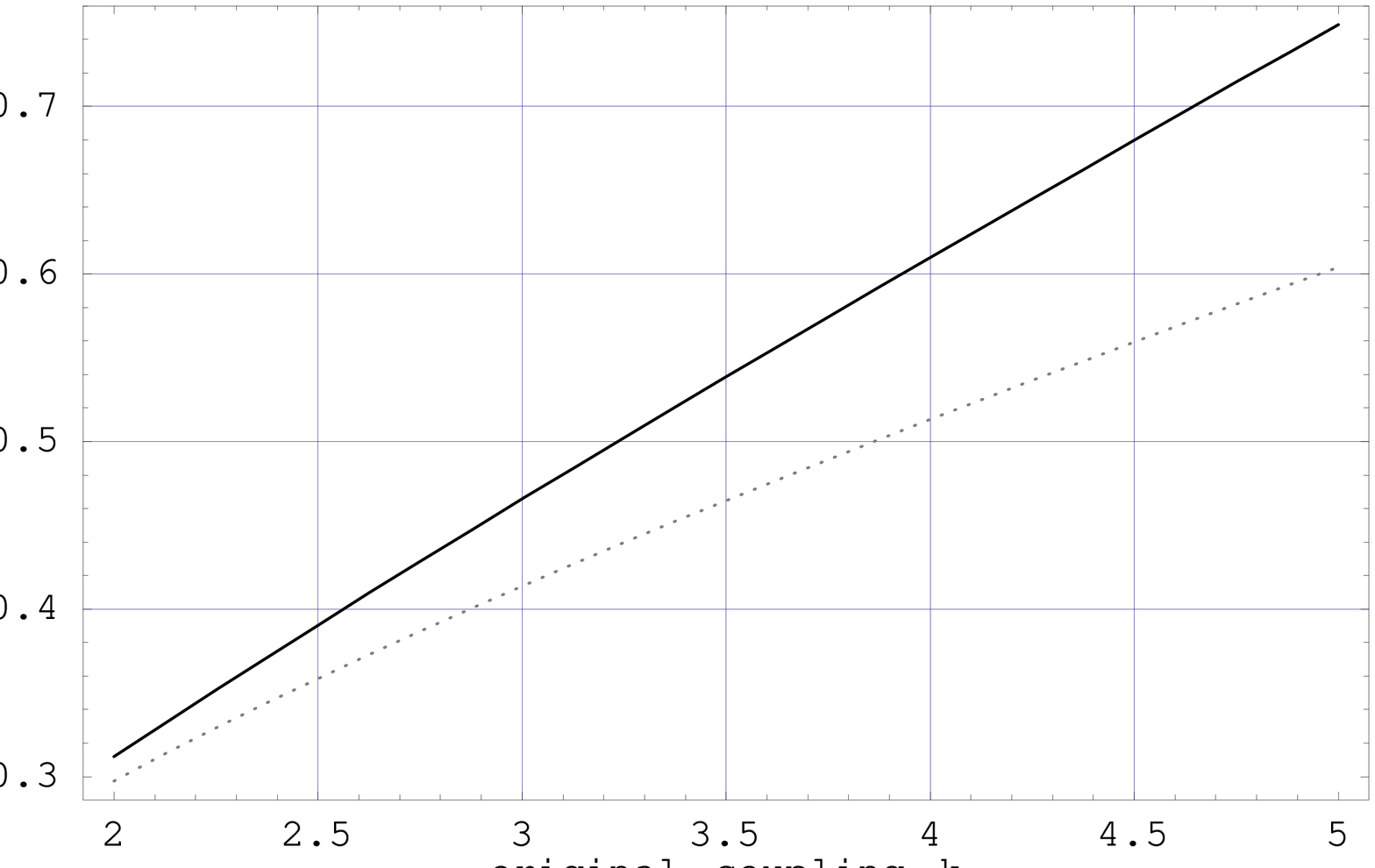,width=8cm,height=5.8cm}}
\vspace{0.8cm}
\caption{\em The RG flow of the enlarged superblock GRRGT (dotted curve)
  in a lower temperature regime compared to the GRRG flow behaviour
  of the superblock (solid curve).}
  \label{twosuper}
\end{figure}

%%%%%%%%%%%%%%%%%%%%%%%%%%%%%%%%%%%%%%%%%%%%%%%%%%%%%
%
To calculate the RG flow behaviour of the constructed GRRGT for the
ferromagnetic isotropic spin-$\frac{1}{2}$ Heisenberg chain we have
to solve (\ref{comm4}) for the effective coupling $K'$. It is convenient
to rewrite the local operators in matrix form and solve the resulting
set of equations. Using (\ref{heisb5}) the matrix representation of
${\cal O}_{{\cal H}'_{I}}\left({\bf K}'\right)$ is given by
\begin{align}
 \label{qtrans4}
 {\cal O}_{{\cal H}'_{I}}\left({\bf K}'\right)
  \;=\;a'\,{\mathbbm {1}}_{2i-1,2i+1}\,+\,b'\,\bsm_{2i-1}\cdot\bsm_{2i+1}
\end{align}
and the coefficients are determined as\\
\begin{align}
 \label{qtrans5}
 a'(K',f')\,& =\,\left[\cosh^3(K')-\sinh^3(K')\right]\,e^{f'}
      \qquad\text{and}\nonumber\\[0.2cm]
  b'(K',f')\,&
      =\,\left[\sinh(K')\cosh^2(K')-\cosh(K')\sinh^2(K')\right]
          \,e^{f'}\,.\nonumber\\
\end{align}
Here we made use of the relation
\begin{align*}
  e^{K'\,(\bsm_{2i-1}\cdot\bsm_{2i+1})\, +f'}
  \,=\,
   e^{K'\,\sm^x_{2i-1}\sm^x_{2i+1}}\,\cdot\, e^{K'\,\sm^y_{2i-1}\sm^y_{2i+1}}
    \,\cdot\, e^{K'\,\sm^z_{2i-1}\sm^z_{2i+1}}\,\cdot\, e^{f'}
\end{align*}
together with a trigonometric expansion using
$(\sm^{\alpha}_{2i-1}\sm^{\alpha}_{2i+1})^2 = {\mathbbm {1}}_{2i-1,2i+1}\,,\;
 \alpha = x,y,z$
and $\sm^x , \sm^y$ and $\sm^z$ as defined in (\ref{heisb5}).
According to (\ref{summarize}) the functional dependence ${\cal O}$ is
invariant under an application of the GRRGT and we define
\begin{align}
 \label{qtrans6}
 G^+_{{\cal H}_{I}\,\otimes\,{\left({\cal H}_{\text{aux}}\right)}_{I}}\,
 {{\cal O}_{{\cal H}_{I}\otimes{\left({\cal H}_{\text{aux}}\right)}_{I}}
   \left({\bf K}\right)}
  \,=\,a(K)\,{\mathbbm{1}}_{2i-1,2i+1}\,+\,b(K)\,\bsm_{2i-1}\cdot\bsm_{2i+1}\,.
\end{align}
In the appendix we prove that this relation is well defined. Inserting
(\ref{qtrans4}) and (\ref{qtrans6}) into the local GRRGT (\ref{comm4})
and solving the resulting set of equations yields
\begin{align}
 \label{qtrans6b}
  a(K)\,=\,a'(K',f')\qquad\text{and}\qquad b(K)\,=\,b'(K',f')\,.
\end{align}
To solve the coupled equations (\ref{qtrans6b}) the parameters $a$ and $b$
need to be calculated. Taking the trace over all the odd sites in
(\ref{qtrans6}) results in
\begin{align}
 \label{qtrans7}
 a(K)\,=\,\frac{1}{4}\text{tr}\,
   \Big\{{{\cal O}_{{\cal H}_{I}\otimes{\left({\cal H}_{\text{aux}}\right)}_{I}}
   \left({\bf K}\right)}\Big\}\,.
\end{align}
From (\ref{summarize}) we conclude the diagonalizability of
${{\cal O}_{{\cal H}_{I}\otimes{\left({\cal H}_{\text{aux}}\right)}_{I}}}$
using a unitary transformation $D\,=\,U^{\dag}\,{H}\,U$ with
$D$ a diagonal matrix. By identifying
$\text{tr}\,\left\{{\cal O}\right\}
     \,=\,\text{tr}\,\left\{\,\exp\left[D\right]\,\right\}$
we explicitly calculate $b(K)$ as
\begin{align}
 \label{Bcalc1}
  b(K)\, &=\,\frac{1}{12}\,\text{tr}\,\left\{\,
     U^{\dag}\,\bsm_{2i-1}\cdot\bsm_{2i+1}\,U\,U^{\dag}\,
          {\cal O}\,U\,\right\}\nonumber\\[0.2cm]
    \, &=\,\frac{1}{12}\,\text{tr}\,\left\{\,
     U^{\dag}\,\bsm_{2i-1}\cdot\bsm_{2i+1}\,U\,
     \exp\left[K\,D_{{\cal H}_{I}\otimes{\left({\cal
     H}_{\text{aux}}\right)}_{I}}\right]\,\right\}\,.\nonumber\\
     \, &=\,\frac{1}{12}\,\text{tr}\,\left\{\,
       U^{\dag}_{{\cal H}_{I}\otimes{\left({\cal H}_{\text{aux}}\right)}_{I}}
       \,H_{{\cal H}'_{I}\otimes{\left({\cal H}'_{\text{aux}}\right)}_{I}}\,
       U\,_{{\cal H}_{I}\otimes{\left({\cal H}_{\text{aux}}\right)}_{I}}
       \exp\left[K\,D_{{\cal H}_{I}
           \otimes{\left({\cal H}_{\text{aux}}\right)}_{I}}
          \right]\,\right\}\,.\nonumber\\
\end{align}
Whereas the computation of the effective parameters $a'$ and $b'$ in
(\ref{qtrans5}) results from a reformulation of (\ref{summarize}) the
calculation of $a$ and $b$ involves the truncation procedure of the
GRRGT. The matrix
$\exp\left[K\,D_{{\cal H}_{I}\otimes{\left({\cal
          H}_{\text{aux}}\right)}_{I}}\right]$
used in the calculation of $a$ and $b$ is a diagonal matrix and the
nonzero elements are the Boltzmann weights
\begin{align}
 \label{Bcalc4}
 \exp\left[K\,D_{{\cal H}_{I}\otimes{\left({\cal
    H}_{\text{aux}}\right)}_{I}}\right]
  \,=\,\exp\left[\frac{1}{k_B\,T}\,E_j\right]
\end{align}
of the superblock
${{\cal H}_{I}\otimes{\left({\cal H}_{\text{aux}}\right)}_{I}}$ where $E_j$
denotes the corresponding energy eigenvalue.\\
The core concepts of the GRRG follow the fundamental ideas of the
density matrix renormalization group (DMRG) as explained in the original
GRRG work \cite{de00}. Relation (\ref{Bcalc4}) defines the density matrix,
originally introduced by R.P. Feynman \cite{fe72}, of the superblock. The
density matrix contains the information about the 'statistical importance'
of each eigenstate of
${{\cal O}_{{\cal H}_{I}\otimes{\left({\cal H}_{\text{aux}}\right)}_{I}}}$
at temperature $T$. Compared to the numerical DMRG
procedure \cite{{wh93},{dl94}} in which only the ground state (target state)
is used for the calculation of the flow behaviour the GRRG uses all
eigenstates weighted by their importance to define the local RGT.
Within a finite temperature RG approach we expect that all the eigenstates
of the superblock will contribute to the RG flow behaviour. The
conventional strategy of using selected eigenvectors for constructing
projection operators $G_{{\cal H}'_{I}}$ and
$G^+_{{\cal H}_{I}\,\otimes\,{\left({\cal H}_{\text{aux}}\right)}_{I}}$,
used to project on a subspace of the total Hilbert space \cite{gmdsv95}, is
therefore not sufficient in a finite temperature approach.\\
For an explicit calculation of the RG flow, we have to solve
(\ref{qtrans6b}) for the effective parameter $K'(K)$. By means of
(\ref{qtrans5}) we obtain
\begin{align}
 \label{Bcalc3}
  K'(K)\,=\,\frac{1}{4}\cdot\ln\left[\frac{a(K)+b(K)}{a(K)-3b(K)}\right]
\end{align}
and a similar expression is obtained for the energy shift  $f'(K)$.
Converting (\ref{Bcalc3}) into an explicit function of the flow
parameter $K$ requires the computation of the traces in (\ref{qtrans7})
and (\ref{Bcalc1}) involving large matrix expressions. The resulting
flow equation $K'(K)$ displays a complicated structure and contains
correction terms with increasing relevance in the low temperature
regime.\\
Figure \ref{ferr} shows a plot of the RG flow for the local
RGT (\ref{comm4}) using (\ref{summarize}) in the ferromagnetic regime,
i.e. $K>0$ and $K'>0$.
% %%%%%%%%%%%%%%%%%%%%%%%%%%%%%%%%%%%%%%%%%%%%%%%%%%%%%%%%%%%%%%%%%
%
% Ferromagnetic superblock plot.
%
% %%%%%%%%%%%%%%%%%%%%%%%%%%%%%%%%%%%%%%%%%%%%%%%%%%%%%%%%%%%%%%%%%
% %%%%%%%%%%%%%%%%%%%%%%%%%%%%%%%%%%%%%%%%%%%%%%%%%%%%%%%%%%%%%%%%%
In figure \ref{ferr} we also plotted the RG flow of the original system
block with no auxiliary space. This flow behaviour is equivalent
to the one calculated by M. Suzuki et al. \cite{st79} valid in the high
temperature limit. The different RG flows deviate from each other except
in the high temperature limit, i.e. $K\rightarrow 0$, where the
correlation block part (\ref{corrdef}) vanishes. From the curves shown
in figure \ref{ferr} we conclude that the high temperature fixed point
regime can be explored without defining an auxiliary space describing
the quantum correlations within the block decomposition.
However, figure \ref{ferr} also illustrates the necessity of including
correlation block terms for computing the RG flow towards lower
temperatures.\\
According to the observed importance of the correlation block part by
approaching lower temperatures, we like to improve the local GRRGT including
higher order correction terms. The special choice of the auxiliary space does
not allow to describe quantum correlations beyond the nearest neighbour sites
of the system block. However, enlarging the auxiliary space by including
the next nearest neighbour (n.n.n.) sites results in the construction of
an enlarged system block. According to the local embedding and truncation
maps (\ref{summarize}) the additional single site spaces can not be marked
as copies within the enlarged superblock.\\
Instead of changing the embedding and truncation maps which in turn demands
for choosing a different invariance relation (\ref{inveq2}), we vary the size
of the original system block, i.e. the {\em scaling} or {\em reduction factor}
$\lambda$ in the GRRGT defined as
\begin{align}
 \label{reduce}
  \lambda\,=\,\frac{\mbox{number of lattice sites in the original spin chain}}
             {\mbox{number of lattice sites in the truncated spin chain}}\;\;.
\end{align}
The previous calculations were based on a one site decimation procedure
equivalent to a local GRRGT with a reduction factor $\lambda =2$.\\
Figure \ref{foursuper} displays two constructions of a superblock for a
system block containing four single site Hilbert spaces with a reduction
factor $\lambda =3$.
% %%%%%%%%%%%%%%%%%%%%%%%%%%%%%%%%%%%%%%%%%%%%%%%%%%%%%%%%%%%%%%%%%
%
% Four site superblock construction.
%
% %%%%%%%%%%%%%%%%%%%%%%%%%%%%%%%%%%%%%%%%%%%%%%%%%%%%%%%%%%%%%%%%%
% %%%%%%%%%%%%%%%%%%%%%%%%%%%%%%%%%%%%%%%%%%%%%%%%%%%%%%%%%%%%%%%%%
In figure \ref{foursuper} a superblock is defined with an
auxiliary space including copies of the n.n. spin sites of the system
block and an {\em enlarged superblock} containing also copies of the
n.n.n. spin sites. According to the geometry of the system block the
n.n. spin sites and the n.n.n. spin sites must be truncated within the
local GRRGT which is consistent with our definition of the embedding and
truncation maps for an active auxiliary space. The enlarged superblock
contains four original system block sites and four further auxiliary
sites.\\
We define a {\em goodness} ${\cal G}$ of the local GRRGT as the ratio of
the number of copies of spin sites contained in the auxiliary space
divided by the number of spin sites within the original system block
\begin{align}
 \label{goodne}
  {\cal G}\,=\,\frac{\mbox{number of copies of spin sites in the auxiliary space}}
             {\mbox{number of spin sites in the system block}}\;\;.
\end{align}
If no auxiliary space is defined ${\cal G}=0$, whereas ${\cal G}>1$ if the
auxiliary space contains more copies of spin sites than original spin sites
are contained in the system block. A sequence of improved local GRRGTs is
generated by enlarging the auxiliary space as visualized for a four site
system block in figure \ref{foursuper}.\\
In figure \ref{fourplot} the RG flow for the superblock and the enlarged
superblock structures depicted in figure \ref{foursuper} are plotted.
% %%%%%%%%%%%%%%%%%%%%%%%%%%%%%%%%%%%%%%%%%%%%%%%%%%%%%%%%%%%%%%%%%
%
% Four site superblock plot.
%
% %%%%%%%%%%%%%%%%%%%%%%%%%%%%%%%%%%%%%%%%%%%%%%%%%%%%%%%%%%%%%%%%%
% %%%%%%%%%%%%%%%%%%%%%%%%%%%%%%%%%%%%%%%%%%%%%%%%%%%%%%%%%%%%%%%%%
Again the dashed curve denotes the RG flow of the original four site
block without any auxiliary space. All different RG flows plotted in
figure \ref{fourplot} show the correct high temperature flow
behaviour by converging to the trivial high temperature fixed point.
Apart from small corrections the superblock GRRGT and the enlarged
superblock GRRGT display the same flow behaviour, indicating that the
auxiliary space constructed by copies of n.n. sites provides a
sufficient description of the quantum correlations in the plotted
regime. However, we expect different flow behaviour for both
superblock GRRGTs at lower temperatures. Figure \ref{twosuper} shows 
the flow behaviour of the local GRRGT constructed from the superblock
and the enlarged superblock displayed in figure \ref{foursuper} away
from the high temperature limit.\\
% %%%%%%%%%%%%%%%%%%%%%%%%%%%%%%%%%%%%%%%%%%%%%%%%%%%%%%%%%%%%%%%%%
%
% Four site two superblocks plot.
%
% %%%%%%%%%%%%%%%%%%%%%%%%%%%%%%%%%%%%%%%%%%%%%%%%%%%%%%%%%%%%%%%%%
% %%%%%%%%%%%%%%%%%%%%%%%%%%%%%%%%%%%%%%%%%%%%%%%%%%%%%%%%%%%%%%%%%
Here we give a comment on the comparison between the GRRG method and
the numerical DMRG procedure invented by S.R.White \cite{wh93} which
provided the core concepts for our work \cite{de00}. The DMRG algorithm
is designed for the numerical calculation of ground state properties
of the physical system and is restricted to $T=0$, i.e. no RG flow
behaviour can be examined \cite{{dl94},{dd95}}. Increasing the size of
the system block in the DMRG method improves the numerical accuracy,
but does not allow for studying the system at finite temperature. Within
a DMRG calculation it is not possible to compute thermodynamic quantities
and examine a phase transition at finite temperature. According to the
concepts of statistical physics \cite{am78} a phase transition at finite
temperature occurs as a nontrivial fixed point in the RG flow whereas the
high and low temperature limits represent trivial fixed points regimes.
The RG flow behaviour in a nontrivial fixed point regime is characterized
by a set of critical exponents defining the type of the phase
transition \cite{am78}.\\
From the GRRG method we calculate the thermal critical exponent
$\nu$ representing the magnetic phase transition using \cite{be91}
\begin{align}
 \label{calcnu}
  \nu \,=\, \frac{\log\left(\frac{\partial K'}{\partial K}{\Big |}_{K^*}\right)}
                 {\log (\lambda)}
\end{align}
where $K^*$ denotes the value of the fixed point coupling in the RG
flow equation (\ref{Bcalc3}).\\
Strongly correlated systems do not exhibit a nontrivial fixed point
in one dimension \cite{mw66}. Solving the RG flow equations plotted
in the foregoing figures therefore leads to the same trivial flow
behaviour as in the classical analogue examined in section \ref{ising}
without exhibiting a nontrivial fixed point. Nontrivial fixed points
occur in quantum spin chains of higher dimensions.\\
In 1975 A.A. Migdal proposed a method for analytical continuation to
higher dimensions of RG recursion formulas for strong coupled systems
exhibiting global symmetries \cite{{mi75a},{mi75b}}. The result of
A.A. Migdal, applicable to a variety of decimation and truncation
procedures \cite{bl82}, was rederived and rigorously analyzed by
L.P. Kadanoff by inventing the Kadanoff bond moving
procedure \cite{ka76}. Both authors assumed a model Hamiltonian with
n.n. interactions. Although L.P. Kadanoff has generalized and corrected
the results of A.A. Migdal, the resulting formula for isotropic quantum
spin models was exactly the same as the equation proposed by A.A. Migdal
given by
\begin{align}
 \label{bondmove}
  K(\lambda L) \,=\, {\lambda}^{d-1}\, R^{\lambda} \left[ K(L) \right] \;.
\end{align}
In relation (\ref{bondmove}) $L$ denotes the lattice constant,
i.e. the distance between two n.n. spin sites and the functional
$R^{\lambda}$ denotes the RGT in the coupling $K$. According to the
notation used by L.P. Kadanoff $\lambda L$ denotes the lattice constant
in the decimated spin chain. The calculations presented in this work
include no explicit dependence on a lattice constant $L$ and we identify
$K'=K(\lambda L)$. We applied the Kadanoff bond moving procedure to the
isotropic XXX spin-$\frac{1}{2}$ Heisenberg model in dimension $d=3$
which exhibits a nontrivial fixed point \cite{st79}. We determined the
nontrivial fixed point for all constructed local GRRGTs and confirmed
that in dimension $d<3$ all GRRG flows exhibit only the trivial fixed
points $K^*=0$ and $K^*=\infty$.\\
In table \ref{table1} we have summarized computed fixed point values
together with the corresponding critical exponents $\nu$ calculated
by equation (\ref{calcnu}) for dimension $d=3$.
% %%%%%%%%%%%%%%%%%%%%%%%%%%%%%%%%%%%%%%%%%%%%%%%%%%%%%%%%%%%%%%%%%
%
% Table of the critical exponents.
%
% %%%%%%%%%%%%%%%%%%%%%%%%%%%%%%%%%%%%%%%%%%%%%%%%%%%%%%%%%%%%%%%%%
\begin{table}[ht]
 \begin{tabular}{l|l|l|l}
  Method & RG flow fixed point $K^*$\; & Critical exponent $\nu$\; &
  Goodness ${\cal G}$\;
  \\[0.1cm] \tableline
  $\lambda =3$ reduction and no auxiliary space
   & $0.522\, (\pm 1.0\cdot 10^{-4})$ & $0.645\, (\pm 1.0\cdot 10^{-4})$ & 0
  \\[0.1cm]
  $\lambda =3$ reduction superblock
   & $0.640\, (\pm 1.0\cdot 10^{-4})$ & $0.470\, (\pm 1.0\cdot 10^{-4})$ & 0.5
  \\[0.1cm]
  $\lambda =3$ reduction enlarged superblock
   & $0.627\, (\pm 1.0\cdot 10^{-4})$ & $0.489\, (\pm 1.0\cdot 10^{-4})$ & 1.0
  \\[0.1cm]
  $\lambda =4$ reduction and no auxiliary space
   & $0.703\, (\pm 1.0\cdot 10^{-4})$ & $0.595\, (\pm 1.0\cdot 10^{-4})$ & 0
  \\[0.1cm]
  $\lambda =4$ reduction superblock
   & $0.851\, (\pm 1.0\cdot 10^{-4})$ & $0.469\, (\pm 1.0\cdot 10^{-4})$ & 0.4
  \\[0.1cm]
  $\lambda =4$ reduction enlarged superblock
   & $0.837\, (\pm 1.0\cdot 10^{-4})$ & $0.475\, (\pm 1.0\cdot 10^{-4})$ & 0.8
  \\[0.1cm]
  Approximate decimation method \cite{st79} & $0.344$ & $0.714$& 0
  \\[0.1cm]
  MFRG combined with decimation \cite{so96}& $0.312$ & $0.758$& -
  \\[0.1cm]
  Mean Field RG (MFRG) \cite{so96}& $0.275$ & $0.450$ & -
  \\
 \end{tabular}
 \vspace{0.4cm}
 \caption{\em The numerical fixed point values and the corresponding
  critical exponents $\nu$ of the isotropic quantum spin-$\frac{1}{2}$
  Heisenberg model in dimension $d=3$ calculated by different methods.}
 \label{table1}
\end{table}
% %%%%%%%%%%%%%%%%%%%%%%%%%%%%%%%%%%%%%%%%%%%%%%%%%%%%%%%%%%%%%%%%%
We applied the GRRGT with a reduction factor $\lambda =3$ and $\lambda =4$
using the superblock and the enlarged superblock structure. The calculated
critical exponents vary between $0.47$ and $0.49$ ordered by the goodness
${\cal G}$ of the approximation used for describing the boundary
conditions. For both reduction factors we furthermore calculated the
critical exponents if no auxiliary space is provided corresponding to an
approximation only valid in the high temperature limit. The calculated
values of the critical exponent deviate significantly from the results
obtained by using the superblock or the enlarged superblock structure.
In table \ref{table1} we further compared the results of the GRRG method
with the outcome of other methods  from the literature. The values of
the critical exponent for the {\em Approximate decimation} method and
the {\em Mean Field RG} method deviate significantly from each other.
Combining both methods resulted in an even higher value for the
critical exponent as compared to the approximate decimation method,
difficult to validate. For strong coupled quantum spin lattices at finite
temperature $T$ no exact approach is available in dimension $d=3$ to
calculate thermodynamic behaviour. The GRRG method with the proposed
auxiliary space is a rigorous and analytic approximation. The
approximation is controlled by a goodness parameter calculated in
equation (\ref{goodne}) yielding consistent results.
%
%%%%%%%%%%%%%%%%%%%%%%%%%%%%%%%%%%%%%%%%%%%%%%%%%%%%%%%%%%%%%%%%%%
\section{The anti-ferromagnetic isotropic Heisenberg chain}
\label{quancorrtanti}
\begin{figure}[ht]
\vspace{0.6cm}
\centerline{
  \psfig{figure=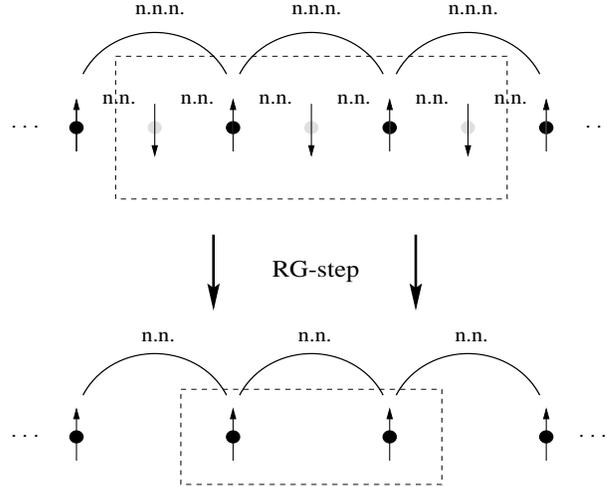,width=8cm,height=6.4cm}}
\vspace{0.8cm}
\caption{\em An application of the RGT for a reduction factor
  $\lambda =2$. The nearest neighbour (n.n.) coupling $K<0$ is
  transformed into an effective n.n. coupling of ferromagnetic
  type $K'>0$.}
  \label{anferrtrans}
\end{figure}

\begin{figure}[ht]
\vspace{0.6cm}
\centerline{
  \psfig{figure=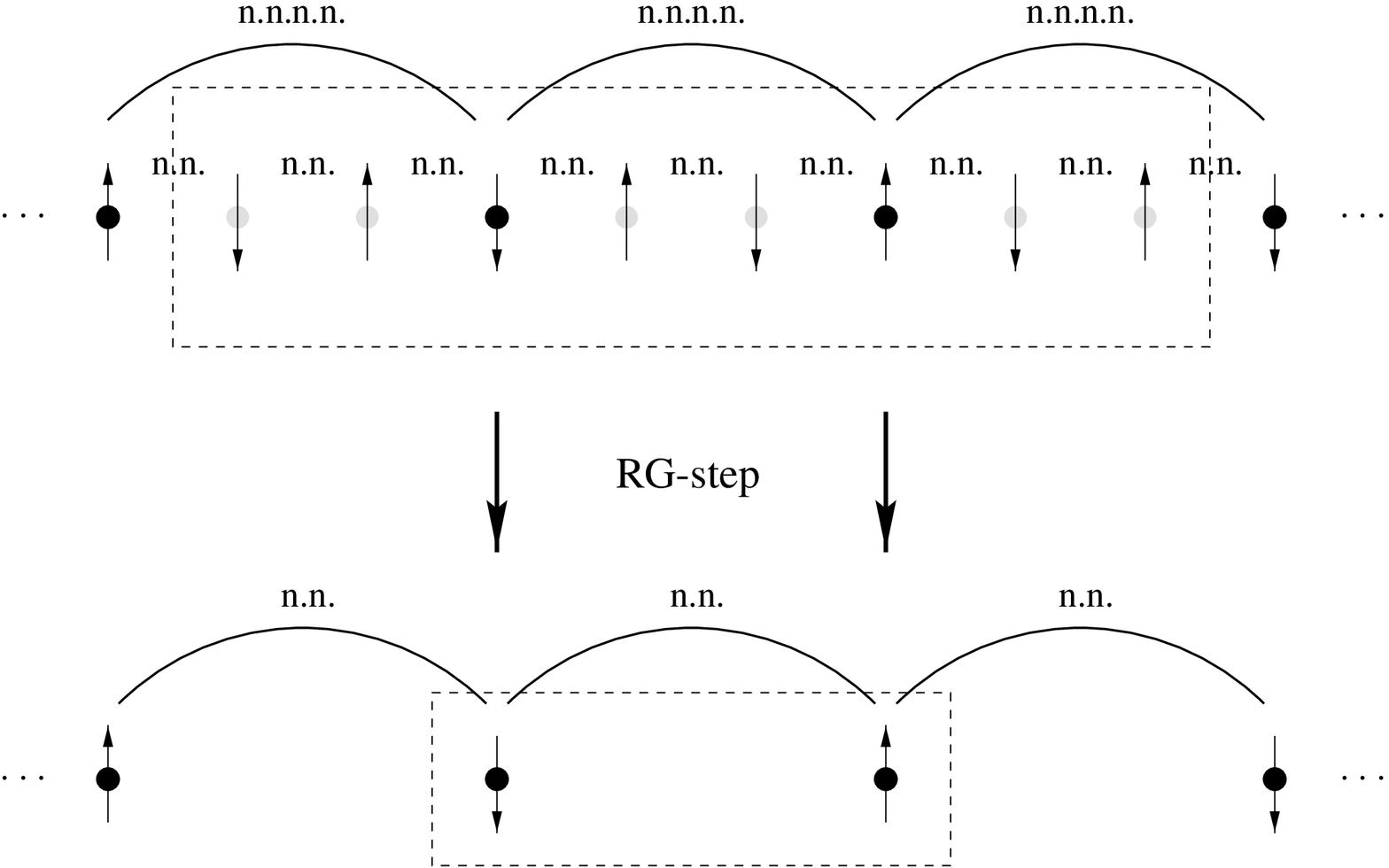,width=9.6cm,height=6.2cm}}
\vspace{0.8cm}
\caption{\em An application of the RGT for a reduction factor
  $\lambda =3$. The anti-ferromagnetic nearest neighbour (n.n.)
  coupling is transformed into an effective n.n. coupling of
  anti-ferromagnetic type $K'<0$.}
  \label{enlanfer}
\end{figure}

\begin{figure}[ht]
\vspace{0.6cm}
\centerline{
  \psfig{figure=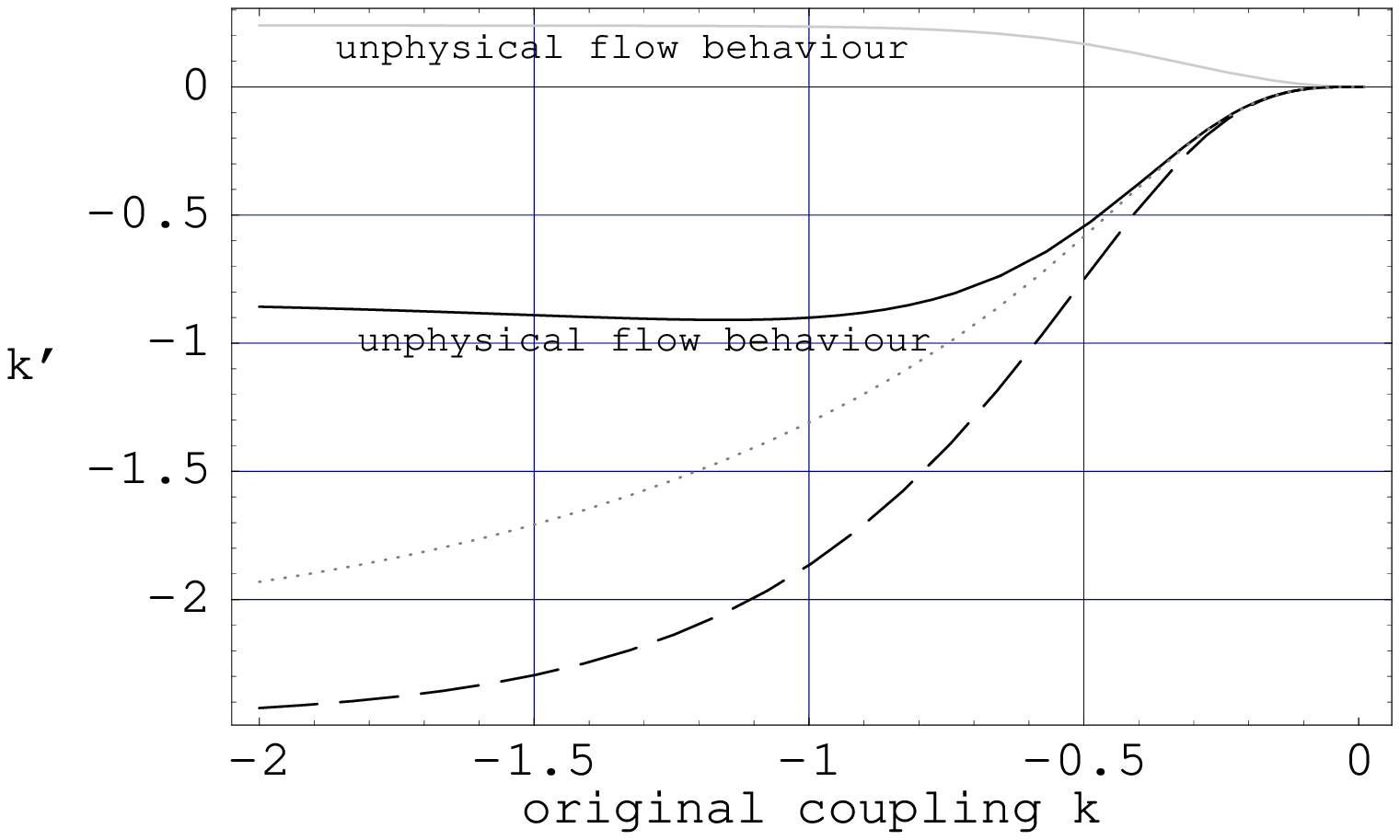,width=9cm,height=6.8cm}}
\vspace{0.8cm}
\caption{\em The lighted and the black solid curve display the
  unphysical GRRG flow behaviour using a superblock structure with a
  reduction factor $\lambda =2$ and $\lambda =3$ respectively. The
  dotted curve depicts the anti-ferromagnetic GRRG flow behaviour
  for the enlarged superblock with a reduction factor $\lambda =3$.
  The dashed curve displays the flow behaviour for the system block
  with a reduction factor $\lambda =3$ using no auxiliary space.}
  \label{anferflow}
\end{figure}

%%%%%%%%%%%%%%%%%%%%%%%%%%%%%%%%%%%%%%%%%%%%%%%%%%%%%%%%%%%%%%%%%%
%
In this section we examine the anti-ferromagnetic regime, i.e. $K<0$.
By using a reduction factor $\lambda =2$ or $\lambda =4$ in the local
GRRGT the anti-ferromagnetic part of the RG flow exhibits an unphysical
behaviour, i. e. applying the local GRRGT once yields a ferromagnetic
coupling $K>0$. The situation is different for a system block structure
with a reduction factor $\lambda =3$. According to the geometry
of the enlarged superblock the GRRG flow shows the correct
anti-ferromagnetic behaviour.\\
Due to the inherent global symmetry of the isotropic quantum
spin-$\frac{1}{2}$ Heisenberg model the eigenstates of the model
Hamiltonian can be represented by the spin $z$-component for each
lattice spin site. Using this representation the ground-state for the
quantum spin-$\frac{1}{2}$ Heisenberg model is represented by an
alternating sequence of {\em spin up} and {\em spin down}
$z$-components. Figure \ref{anferrtrans} visualizes a RG-step, i.e.
applying the RGT once, using a reduction factor $\lambda =2$.
% %%%%%%%%%%%%%%%%%%%%%%%%%%%%%%%%%%%%%%%%%%%%%%%%%%%%%%%%%%%%%%%%%
%
% Anti-ferromagnetic superblock transformation.
%
% %%%%%%%%%%%%%%%%%%%%%%%%%%%%%%%%%%%%%%%%%%%%%%%%%%%%%%%%%%%%%%%%%
% %%%%%%%%%%%%%%%%%%%%%%%%%%%%%%%%%%%%%%%%%%%%%%%%%%%%%%%%%%%%%%%%%
The original n.n. coupling of anti-ferromagnetic type is transformed
into a n.n. coupling of ferromagnetic type according to the structure
of the system block. By using a reduction factor $\lambda =2$ the
geometry of the spin lattice does not allow for calculating an
anti-ferromagnetic GRRG flow behaviour.\\
By choosing a reduction factor $\lambda =3$ the geometry of the system
block changes. In figure \ref{enlanfer} we depict a RG-step for a
system block composed of four sites.
% %%%%%%%%%%%%%%%%%%%%%%%%%%%%%%%%%%%%%%%%%%%%%%%%%%%%%%%%%%%%%%%%%
%
% Anti-ferromagnetic enlarged superblock transformation.
%
% %%%%%%%%%%%%%%%%%%%%%%%%%%%%%%%%%%%%%%%%%%%%%%%%%%%%%%%%%%%%%%%%%
% %%%%%%%%%%%%%%%%%%%%%%%%%%%%%%%%%%%%%%%%%%%%%%%%%%%%%%%%%%%%%%%%%
By applying the RGT the original anti-ferromagnetic n.n. coupling
$K<0$ is transformed into an effective anti-ferromagnetic n.n.
coupling $K'<0$. According to the structure of the system block it is
possible to construct an entire anti-ferromagnetic GRRG flow using the
enlarged superblock structure.\\
In figure \ref{anferflow} the anti-ferromagnetic GRRG flow using the
enlarged superblock structure depicted in figure \ref{enlanfer} is
plotted by the dotted curve.
% %%%%%%%%%%%%%%%%%%%%%%%%%%%%%%%%%%%%%%%%%%%%%%%%%%%%%%%%%%%%%%%%%
%
% Four site two superblocks plot.
%
% %%%%%%%%%%%%%%%%%%%%%%%%%%%%%%%%%%%%%%%%%%%%%%%%%%%%%%%%%%%%%%%%%
% %%%%%%%%%%%%%%%%%%%%%%%%%%%%%%%%%%%%%%%%%%%%%%%%%%%%%%%%%%%%%%%%%
The dashed curve displays the flow behaviour using a system block
containing four spin sites without an auxiliary space. Although the
RGT respects the geometry of an anti-ferromagnetic quantum spin
chain no quantum block correlations are taken into account which is
valid in the high temperature limit $T\longrightarrow \infty$. In
figure \ref{anferflow} two further GRRG flows were plotted using a
superblock structure with a reduction factor $\lambda =2$ and
$\lambda =3$. The lighted solid curve shows the expected unphysical
behaviour for the reduction factor $\lambda =2$ as depicted in figure
\ref{anferrtrans}. The black solid curve displays the GRRG flow
according to the superblock with a reduction factor $\lambda =3$
depicted in figure \ref{foursuper}. Although the system block,
containing four original spin sites, respects the anti-ferromagnetic
geometry, the auxiliary space does not. However, in the high
temperature regime the GRRG flow for the superblock and the enlarged
superblock structure using a reduction factor $\lambda =3$ display
equal behaviour. In the finite temperature regime ground state
properties become increasingly important and the auxiliary space
for the superblock with a reduction factor $\lambda =3$ does not
provide the correct anti-ferromagnetic boundary conditions.\\
%
%%%%%%%%%%%%%%%%%%%%%%%%%%%%%%%%%%%%%%%%%%%%%%%%%%%%%
\section{Conclusions}
%%%%%%%%%%%%%%%%%%%%%%%%%%%%%%%%%%%%%%%%%%%%%%%%%%%%%
%
In this paper we applied the Generalized Real-space Renormalization
Group (GRRG) method to the isotropic spin-$\frac{1}{2}$ Heisenberg model.
The analytic method is non-perturbative and yields an approximate RG flow
behaviour in the finite temperature regime controlled by a quality measure,
i.e. the goodness parameter.\\
In section \ref{quancorrt} we constructed a local GRRG transformation
based on a chosen auxiliary space. The spin site decimation was
formulated in terms of an embedding and truncation map.\\
In section \ref{quanflowt} we examined the ferromagnetic RG flow
behaviour for the superblock and the enlarged superblock structure
using different reduction factors for the spin site decimation. We
explored the RG flow behaviour in three dimensions using the Kadanoff
bond moving procedure originally introduced by A.A. Migdal as a method
for analytical continuation to higher dimensions in RG recursion
formulas. The nontrivial finite temperature fixed point of the 
different RG flows was calculated and the related critical exponent was
examined according to the goodness of the auxiliary space approximation.
The critical exponents computed by the GRRG method were compared to other
results from the literature, mostly calculated by numerical techniques
without providing a quality measure of the approximation.\\
In section \ref{quancorrtanti} the anti-ferromagnetic part of the flow
behaviour was explored. Only system block structures with an odd
reduction factor allowed for a local GRRG transformation respecting the
anti-ferromagnetic geometry of the quantum spin chain. The enlarged
superblock GRRG flow for a two site decimation in the quantum spin
chain exhibited the correct physical flow behaviour and was proved as
a valid approximation in the finite temperature regime. The auxiliary
space of the superblock for the two site decimation did not provide the
correct description of the anti-ferromagnetic boundary conditions
resulted in an unphysical flow behaviour in the lower temperature
regime.\\
Applications of the GRRG method to other quantum systems such as the
Hubbard model will be reported in the future, although further
development of auxiliary spaces is the core concept of the GRRG approach.
In this work we showed that away from the high temperature fixed point
regime boundary conditions become increasingly important in the
calculation of the RG flow behaviour. An exact GRRG method provides
all possible boundary conditions at an arbitrary temperature. We
therefore have constructed an exact local GRRG transformation for the
spin-$\frac{1}{2}$ Heisenberg model using a passive auxiliary space
providing all necessary boundary conditions between adjacent system
blocks \cite{de01}.\\
Although the GRRG method is designed as an analytic RG approach we are
working on a numerical implementation of a GRRG algorithm applicable
to nonlinear partial differential equations. A detailed description of
the mathematical formulation and the numerical implementation of the
algorithm will be presented in the near future. We compute critical
exponents to determine the universality class of the nonlinear partial
differential equation with a reduced number of degrees of
freedom.\\[0.8cm]
%
%%%%%%%%%%%%%%%%%%%%%%%%%%%%%%%%%%%%%%%%%%%%%%%%%%%%%
{\leftline{\large\bf Acknowledgment}}\\[0.2cm]
The author thanks Abdollah Langari for introducing him to the Migdal Kadanoff
RG and for the initial help in this work. The author further sincerely
thanks Prof. P. Stichel for useful comments on the physical content of the
manuscript. For encouraging discussions the author likes to give special thanks
to Javier Rodriguez Laguna, Prof. A. Kl\"umper and Reiner Raupach. The author
thanks the anonymous reviewer for the detailed studying of the manuscript and
the helpful suggestions.\\[0.8cm]
%
%%%%%%%%%%%%%%%%%%%%%%%%%%%%%%%%%%%%%%%%%%%%%%%%%%%%%
\centerline{\large\bf Appendix}\\[0.4cm]
%%%%%%%%%%%%%%%%%%%%%%%%%%%%%%%%%%%%%%%%%%%%%%%%%%%%%
%
In order to derive equation (\ref{qtrans6}) we introduce the abbreviations
\begin{align*}
\text{tr}_{\text{even}}\,\left\{\,
 \exp\left[K\,H_{{\cal H}_{I}\otimes{\left({\cal
     H}_{\text{aux}}\right)}_{I}}\right]\,\right\}
 \;=\; \text{tr}_{\text{even}}\,\left\{\,\varphi(\{\bsm_i\}_{i\in I})\,\right\}
  \;=\; \chi\left(\bsm_{2i-1},\bsm_{2i+1}\right)
\end{align*}
and $\varphi$ and $\chi$ are functions of the Pauli spin matrices
defined in (\ref{heisb5}). Using the rotation symmetry we derive an
equivalent representation $\{\bsm'_i\}$ by
\begin{align*}
 \varphi(\{\bsm_i\})\;=\;\varphi(\{\bsm'_i\})
\end{align*}
with the rotation map $R^{\text{rot}}$ defined as
\begin{align*}
 (\sm_i)'_j\;=\; R^{\text{rot}}_{j k}\, (\sm_i)_k\;.
\end{align*}
From
\begin{align}
 \chi\left(\bsm_{2i-1},\bsm_{2i+1}\right)
 \;=\; &\text{tr}_{\text{even}}\,\Big\{\,\varphi(\{\bsm_i\})\,\Big\}
 \nonumber\\[0.2cm]
  \;=\; &\text{tr}_{\text{even}'}\,\Big\{\,\varphi(\{\bsm_i\})\,\Big\}
 \nonumber\\[0.2cm]
  \;=\; &\text{tr}_{\text{even}'}\,\Big\{\,\varphi(\{\bsm'_i\})\,\Big\}
 \nonumber\\[0.2cm]
  \;=\; &\chi\left(\bsm'_{2i-1},\bsm'_{2i+1}\right)\nonumber
\end{align}
we deduce
\begin{align*}
 \chi\left(\bsm_{2i-1},\bsm_{2i+1}\right)
  \;=\;\chi\left(\bsm_{2i-1}\,\cdot\,\bsm_{2i+1}\right)\;.
\end{align*}
Using (\ref{qtrans4}) we derive (\ref{qtrans6}).\\[0.1cm]
%
%
%%%%%%%%%%%%%%%%%%%%%%%%%%%%%%%%%%%%%%%%%%%%%%%%%%%%%


\begin{thebibliography}{99}
%
\vspace*{-0.2cm}
\centerline{\large\bf References}
\vspace*{0.4cm}
%%%%%%%%%%%%%%%%%%%%%%%%%%%%%%%%%%%%%%%%%%%%%%%%%%%%%
%
%
\bibitem{ka66}
{\sc L.P. Kadanoff}.
\newblock {\em Physics 2, 263}  (1966).
%
\bibitem{wi71}
{\sc K.G. Wilson}.
\newblock {\em Phys. Rev. B4, 3174}  (1971).
\newblock {\em Phys. Rev. B4, 3184}  (1971).
\newblock {\em Phys. Rev. Lett. 28, 548}  (1972).
%
\bibitem{gmdsv95}
{\sc J. Gonzalez}, {\sc M. A. Martin-Delgado}, {\sc G. Sierra}
 and {\sc A.H. Vozmediano}.
\newblock In: {\em Quantum Electron Liquids and High-$T_c$ Superconductivity}.
\newblock {\em Lecture Notes in Physics, m38, Springer, Berlin}  (1995).
%
\bibitem{so96}
{\sc J.R. de Sousa}.
\newblock {\em Phys. Lett. A216, 321}  (1996).
%
\bibitem{gmh98}
{\sc N.D. Goldenfeld}, {\sc A. McKane} and {\sc Q. Hou}.
\newblock {\em J. Stat. Phys. 93, 699}  (1998).
%
\bibitem{cmp98}
{\sc C. Castellano}, {\sc M. Marsili} and {\sc L. Pietronero}.
\newblock {\em Phys. Rev. Lett. 80, 3527}  (1998).
%
\bibitem{de00}
{\sc A. Degenhard}.
\newblock {\em J. Phys. A: Math. Gen. 33 No 35, 6173}  (2000).
%
\bibitem{ka76}
{\sc L.P. Kadanoff}.
\newblock {\em Ann. Phys. (N.Y.) 100, 359}  (1976).
%
\bibitem{hk73}
{\sc A. Houghton} and {\sc L.P. Kadanoff}.
\newblock In: {\em Proceedings of 1973 Temple University Conference on
  Critical Phenomena and Quantum Field Theory}.
\newblock {\em Department of Physics, Temple University}  (1973).
%
\bibitem{mi76a}
{\sc A.A. Migdal}.
\newblock {\em Sov. Phys. JETP 42, 413}  (1976).
%
\bibitem{mi76b}
{\sc A.A. Migdal}.
\newblock {\em Sov. Phys. JETP 42, 743}  (1976).
%
\bibitem{hk75}
{\sc A. Houghton} and {\sc L.P. Kadanoff}.
\newblock {\em Phys. Rev. B11, 377}  (1975).
%
\bibitem{mw66}
{\sc N. Mermin} and {\sc H. Wagner}.
\newblock {\em Phys. Rev. Lett. 17, 1133}  (1966).
%
\bibitem{he28}
{\sc W. Heisenberg}
\newblock {\em Z. Physik 49, 619}  (1928).
%
\bibitem{bl30}
{\sc F. Bloch}.
\newblock {\em Z. Physik 61, 206}  (1930).
%
\bibitem{su76}
{\sc M. Suzuki}.
\newblock {\em Comm. Math. Phys. 51, 183}  (1976).
%
\bibitem{st79}
{\sc M. Suzuki} and {\sc H. Takano}.
\newblock {\em Phys. Lett. 69, 426}  (1979).
%
\bibitem{de01}
{\sc A. Degenhard}.
\newblock An algebraic general RG approach.
\newblock {\em To be submitted.} (2001).
%
\bibitem{fe72}
{\sc R.P. Feynman}.
\newblock Statistical Mechanics: A Set of Lectures.
\newblock {\em Benjamin, Reading, MA}  (1972).
%
\bibitem{wh93}
{\sc S.R. White}.
\newblock {\em Phys. Rev. B48, 10345}  (1993).
%
\bibitem{dl94}

{\sc A. Drzewi$\acute {\sc n}$ski} and {\sc J.M.J. van Leeuwen}.
\newblock {\em Phys. Rev. B49, 403}  (1994).
%
\bibitem{dd95}
{\sc A. Drzewi$\acute {\sc n}$ski} and {\sc R. Dekeyser}.
\newblock {\em Phys. Rev. B51, 15218}  (1995).
%
\bibitem{am78}
{\sc D.J. Amit}.
\newblock Field Theory, the Renormalization Group,
          and Critical Phenomena.
\newblock {\em McGraw-Hill, Inc.}  (1978).
%
\bibitem{be91}
{\sc M. le Bellac}.
\newblock Quantum and Statistical Field Theory.
\newblock {\em Clarendon Press, Oxford}  (1991).
%
\bibitem{mi75a}
{\sc A.A. Migdal}.
\newblock {\em Z. Eksper. Teoret. Fiz. 69, 810}  (1975).
%
\bibitem{mi75b}
{\sc A.A. Migdal}.
\newblock {\em Z. Eksper. Teoret. Fiz. 69, 1457}  (1975).
%
\bibitem{bl82}
{\sc T.W. Burkhardt} and {\sc J.M.J. van Leeuwen}.
\newblock Real-Space Renormalization.
\newblock {\em Topics in Current Physics, Springer, 30, Berlin}  (1982).
%
%
\end{thebibliography}
\end{document}